\documentclass[aps,pre,twocolumn,floatfix,superscriptaddress,showpacs,showkeys]{revtex4-2}
\usepackage{epsf,amsmath,amssymb,verbatim,color,multirow,pifont,mathrsfs,soul,amsthm, wasysym}
\usepackage{graphicx, upgreek}
\usepackage[us,12hr]{datetime}

\begin{document}

\title{Discontinuous transition in explosive percolation via local suppression}
\author{Young Sul Cho}
\email{yscho@jbnu.ac.kr}
\affiliation{Department of Physics, Jeonbuk National University, Jeonju 54896, Rep. of Korea}
\affiliation{Research Institute of Physics and Chemistry, Jeonbuk National University, Jeonju 54896, Rep. of Korea}

\date{\today}

\begin{abstract} 
We study an explosive percolation model in which a link is randomly added and neighboring nodes sequentially rewire their links to suppress the growth of large clusters. In this manner, the rewiring nodes spread outward starting from the initial node closest to the added link. We show that a discontinuous transition emerges even when the total number of rewiring nodes after each link addition is finite. This finding implies that adding a link using the information of the cluster sizes attached to a finite set of link candidates (local information) can lead to a discontinuous transition if link rewiring is allowed. This result thus extends the previous result that a discontinuous transition arises only when a link is added using the information of the cluster sizes attached to an infinite number of link candidates (global information) in the absence of rewiring.
\end{abstract}

\maketitle

\section{Introduction}

The discontinuous percolation transition (DPT) is a phenomenon in which a fraction of nodes in a macroscopic cluster jumps discontinuously as the link occupation fraction crosses a threshold~\cite{explosive_phenomena}. Various theoretical models have been proposed to explain the DPT, such as k-core (bootstrap) percolation~\cite{Bootstrap, kcore, kcore3, lee2016kcore}, percolation in interdependent networks~\cite{InterdependentHavlin, InterdependentDorogovtsev, li2012interdependentlattice, grassberger2015interdependentlattice}, and others~\cite{triadic2023, yang2012core, minsuk2024shortest}. In these models, cascades of link deletions induce a DPT in k-core and interdependent networks when the initial link occupation fraction falls below the threshold~\cite{lee2016hybrid}.

We study the explosive percolation (EP) model in an attempt to reveal a DPT through the sequential occupation of links without the deletion of occupied links~\cite{Achlioptas:2009, souza_nphy}.
In an EP model, at each time step, the optimal link that suppresses the growth of large clusters is occupied using information from a set of link candidates, where this information is referred to as local when the set is finite and global when it is infinite. As a result, an abrupt percolation transition occurs as the link occupation fraction $p$ exceeds the threshold. But although the transition appears abrupt, it is continuous when local information is used~\cite{makse, ziff_prl, filippo_pre:2010, Jan_gap2, tricritical, hklee, choi2011site, dacosta_prl, grassberger, moon2025thermodynamics}, and becomes discontinuous only when global information is used~\cite{riordan, cho_science, Half_restricted, smohdicpt_2018, largest, resource, local2018}.

In such an EP model, each node is not allowed to rewire its links, even when rewiring could reduce its cluster size, as is often adopted in various real systems such as a quarantine~\cite{quaratine}.
Recently, it was reported that a DPT emerges when nodes consistently rewire their links to reduce their cluster sizes until a steady state is reached for each $p$~\cite{yschosciadv2025, yschodiscPTtree2025}.
This result implies that a DPT emerges when each node rewires its links using local information of its neighbors, which indeed overcomes the limitation of the previous EP model that requires global information for a DPT after each link addition by incorporating link rewiring. However, it remains unclear whether a dynamical process can lead to a DPT, like the EP model where a link is added and a finite number of nodes rewire.

\begin{figure}[t!]
\includegraphics[width=0.9\linewidth]{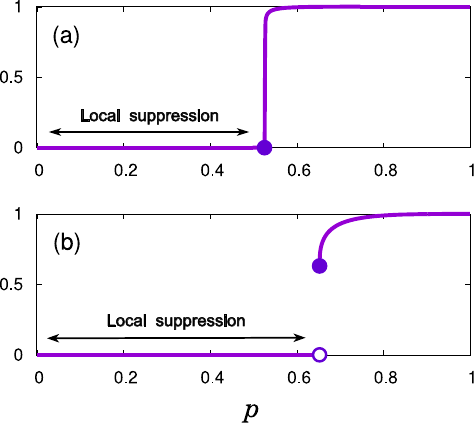}
\caption{Schematic diagram of the fraction of nodes in a macroscopic cluster in the thermodynamic limit $N \rightarrow \infty$ for (a) a previous explosive percolation (EP) model and (b) our EP model. Solid circles denote the values at $p=p_c$, while the open circle indicates the value approached as $p$ approaches $p_c$ from below.}
\label{Fig:PinfSchematic}
\end{figure}

In this paper, we introduce an EP model in which a link is randomly occupied and neighboring nodes sequentially rewire their links starting from the node closest to the occupied link to suppress the growth of large clusters.
In this manner, the size of the largest cluster reaches a steady state after each link occupation, leading to the emergence of a DPT.
We consider this EP model on two types of networks: a Bethe lattice branch and a bipartite network. On the branch, we show that the number of rewiring nodes remains finite up to the threshold. On the bipartite network, our simulations indicate that the number of rewiring nodes is finite below the threshold but diverges as the threshold is approached. We provide a brief explanation of this result.

We also note here that in the previous EP model, an infinite number of link candidates is required to reveal a DPT even below the threshold~\cite{riordan}. Therefore, in the thermodynamic limit $N \rightarrow \infty$, such EP models require global information at each $p$ before the threshold to exhibit a DPT. In contrast, our EP model reveals a DPT using only local information for each $p$ before the threshold. This advance is schematically illustrated in Fig.~\ref{Fig:PinfSchematic}.

This paper is organized as follows. In Sec.~\ref{sec:branch} we introduce the result of the EP model on the branch, and in Sec.~\ref{sec:bipartite} we introduce the result of the EP model on the bipartite network. Then in Sec.~\ref{sec:discussion}, we discuss how our results extend the findings of previous EP models. Supporting data are provided in Appendix A and B.

\section{Explosive percolation in a Bethe lattice branch}
\label{sec:branch}

\begin{figure}[t!]
\includegraphics[width=0.8\linewidth]{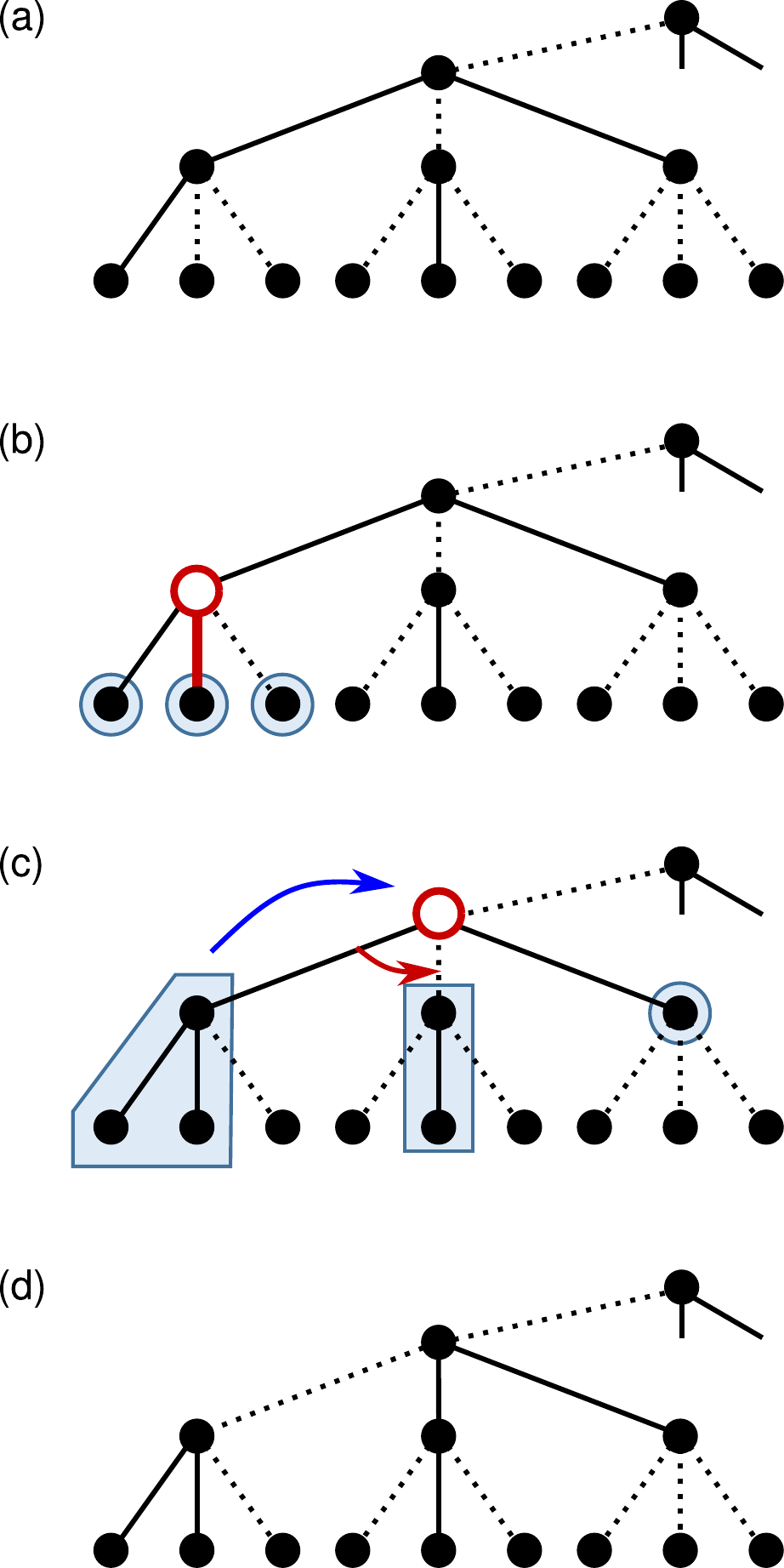}
\caption{Schematic diagram of one link addition in the EP model on a branch with $L=4$. 
(a) Initial configuration, where solid lines denote occupied links and dotted lines denote unoccupied links.
(b) One link (thick red solid line) is added to the initial rewiring node (empty circle) following steps (b1) and (b2) given in Sec.~\ref{sec:branch}.
(c) The node (empty circle) reached via the occupied link from below rewires its links in ascending order of neighboring cluster sizes.
(d) The link to the parent of the rewiring node in (c) is unoccupied, and thus the process of adding one link terminates.}
\label{Fig:SchematicBranch}
\end{figure}

We study the EP model on a branch of the Bethe lattice~\cite{saberi_bethe}, the latter of which is a tree composed of nodes with a constant degree $z=4$. To implement a branch of finite size, we define a branch with $L$ layers as follows. In each layer indexed by $\ell$ $(0 \leq \ell \leq L-1)$, $(z-1)^{\ell}$ nodes exist. For $0 \leq \ell \leq L-2$, each node in the $\ell$th layer attaches links to its $(z-1)$ neighbors in the $(\ell+1)$th layer. Then, the total number of nodes is $N = [(z-1)^{L} - 1]/(z-2)$, and the total number of links is $N-1$, since the branch is also a tree.

Each link is either occupied or unoccupied. A cluster is defined as a set of nodes that are mutually reachable via occupied links.

In a previous study~\cite{yschosciadv2025}, we showed that a DPT emerges when each node rewires its links to suppress the growth of large clusters on a branch. 
Specifically, $p(N-1)$ number of links are initially occupied randomly on the branch. At each time step, a node is selected randomly, where $n$ denotes the number of occupied links of the node. The node rewires its links to neighbors in the lower layer (larger $\ell$) according to the following steps:
\begin{itemize}
\item[(a1)] It disconnects all of its occupied links. 
\item[(a2)] It evaluates the cluster sizes of its neighbors in the lower layer.
\item[(a3)] It occupies $n$ links to neighbors in the lower layer in ascending order of their cluster sizes.
\end{itemize}
In short, each node rewires its links to neighbors in the lower layer in ascending order of their cluster sizes.
As a result, the growth of large clusters is suppressed by the rewiring of each node.
This process continues until a steady state is reached in which each node is connected to its neighbors in ascending order of their cluster sizes.

The order parameter is given by $P_{\infty}(p)$, which is the probability that the root $(\ell=0)$ belongs to the spanning cluster,
where the spanning cluster includes at least one node in the outermost layer $(\ell=L-1)$.
$P_{\infty}(p)$ of the steady state exhibits a DPT at $p=p_c$, such that $P_{\infty}(p)>0$ $(=0)$ for $p \geq p_c$ $(< p_c)$. The analytical derivation of $P_{\infty}(p)$ is summarized in Appendix A.

We now revise this model to reflect a dynamical process where nodes in the EP model rewire their links to reach a steady state after a link is added. 
In this manner, each link is added in a steady state, and thus only a small number of rewiring nodes is sufficient to reach a steady state after a link is added. 
We show below that $P_{\infty}(p)$ in this EP model coincides with that of the steady state, confirming that the EP model indeed exhibits a DPT with only a small number of rewiring nodes for each $p$. 
We note that the total number of rewiring nodes for each value of $p$ is extensive to $N$ in the previous study, as the rewiring process begins from a randomly occupied network that does not correspond to a steady state.

We introduce the dynamical rule of the EP model on the branch as follows. At each time step $p \rightarrow p + 1/(N-1)$, one link is added according to the following steps:
\begin{itemize}
\item[(b1)] One unoccupied link is randomly selected and then occupied. The node in the upper layer (smaller $\ell$) among the two ends of the occupied link becomes the initial rewiring node.  
\item[(b2)] The rewiring node rewires its links to neighbors in the lower layer in ascending order of cluster sizes, following steps (a1)--(a3).
\item[(b3)] If the rewiring node is connected to its parent (neighbor in the upper layer), the parent becomes the rewiring node, and step (b2) is repeated; otherwise, the process terminates.
\end{itemize}
A schematic diagram of this process is shown in Fig.~\ref{Fig:SchematicBranch}.

\begin{figure}[t!]
\includegraphics[width=0.8\linewidth]{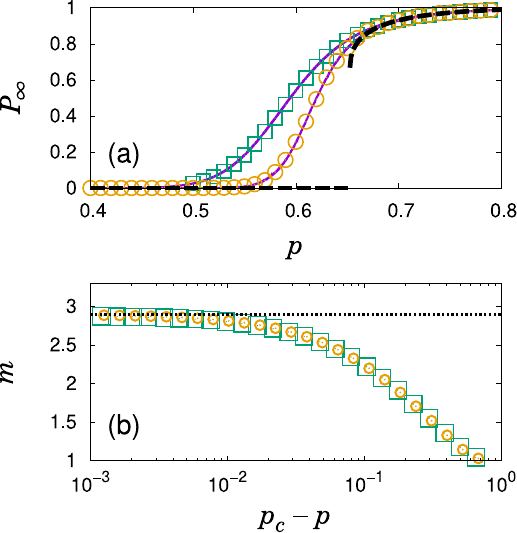}
\caption{(a) $P_{\infty}(p)$ of the EP model (solid line) and the steady state (symbols) on the branch with $L=10$ ($\square$ with line) and $16$ ($\bigcirc$ with line).     
The dashed line is the theoretical curve of $P_{\infty}(p)$
in the thermodynamic limit $L \rightarrow \infty$, as described in Appendix A.
(b) $m$ vs. $p_c-p$ on the branch with $L = 10$ ($\square$) and $16$ $(\circ)$. The dotted line indicates the estimated value $m(p_c) = 2.894$.
}
\label{Fig:BranchPinf_FracRewire}
\end{figure}

As shown in Fig.~\ref{Fig:BranchPinf_FracRewire}(a), $P_{\infty}(p)$ of the EP model coincides with that of the steady state for the same $L$.
We discuss why steps (b1)--(b3) preserve the steady state, in which every node connects to its neighbors in the lower layer in ascending order of their cluster sizes.

A link is added, and the parent of the link becomes the initial rewiring node according to step (b1). The cluster size of the rewiring node increases initially due to the added link. This increase does not cause nodes in the lower layer to rewire, since each node considers only the cluster sizes in the lower layer. Therefore, it is sufficient to examine only the parent of the rewiring node. 
If the rewiring node is connected to its parent, the increase in its cluster size may cause the parent to rewire its link to another node. To ensure the steady state, the parent must itself be treated as a rewiring node and updated according to step (b2). Furthermore, the cluster size of the parent also increases via the link addition, since the newly connected neighbor through link rewiring belongs to a larger cluster than the originally connected one.
In summary, the parent becomes a rewiring node, and its cluster size increases through the link addition. This situation is equivalent to that in the beginning of this paragraph, with the parent treated as the rewiring node. Therefore, steps (b2) and (b3) should be repeated until a rewiring node is disconnected from its parent.

If a rewiring node is disconnected from its parent, the parent is guaranteed not to rewire, since it is already connected to its neighbors in the lower layer in ascending order of their cluster sizes, in which case only the size of the disconnected cluster increases through the link addition. The cluster size of the parent therefore does not change and no further rewiring occurs. Consequently, the rewiring process triggered by the single link addition terminates.

In Fig.~\ref{Fig:BranchPinf_FracRewire}(b), we find that the number of rewiring nodes denoted by $m(p)$ increases monotonically and remains finite up to $p_c$. We briefly argue why $m(p_c)$ does not diverge. Following steps (b1)--(b3), $m$ is equivalent to the length of the sequence of connected nodes in the upward direction (toward smaller $\ell$), starting from the initial rewiring node. If the average value of $m$ diverges at $p_c$, the cluster containing the root must include a finite fraction of nodes at $p_c$ for a given finite $L$. However, this contradicts the property that the macroscopic cluster at $p_c$ is an infinite branching tree, which does not contain a finite fraction of nodes. Accordingly, $m$ must remain finite at $p_c$.

\section{Explosive percolation in a Bipartite network}
\label{sec:bipartite}

Next we consider a bipartite network composed of two partitions, each containing $N/2$ nodes. Each node $i$ in the first partition $(1 \leq i \leq N/2)$ attaches links to three randomly selected neighbors $j$ in the second partition $(N/2+1 \leq j \leq N)$. 
Then, the total number of links is given by $3N/2$.
Each link is either occupied or unoccupied, and a cluster is defined as a set of nodes that are mutually reachable through occupied links.

In a previous study~\cite{yschodiscPTtree2025}, we showed that a DPT emerges when each node in the first partition rewires its links to suppress
the growth of large clusters in the bipartite network. Here, nodes in the second partition are intentionally not allowed to rewire, in order to preserve the number of occupied links attached to each rewiring node in the first partition and to make the process analytically tractable.

Specifically, $p(3N/2)$ number of links are initially occupied randomly in the bipartite network. At each time step, a node in the first partition is selected randomly, where $n$ denotes the number of occupied links of the node. The node rewires its links to neighbors according to the following steps:
\begin{itemize}
\item[(c1)] It disconnects all of its occupied links. 
\item[(c2)] It evaluates the cluster sizes of its neighbors.
\item[(c3)] It occupies $n$ links to its neighbors in ascending order of their cluster sizes.
\end{itemize}
In short, each node in the first partition rewires its links to neighbors in ascending order of their cluster sizes. As a result, the growth of large clusters is suppressed by the rewiring of each node in the first partition. This process continues until a steady state is reached.

\begin{figure}[t!]
\includegraphics[width=1.0\linewidth]{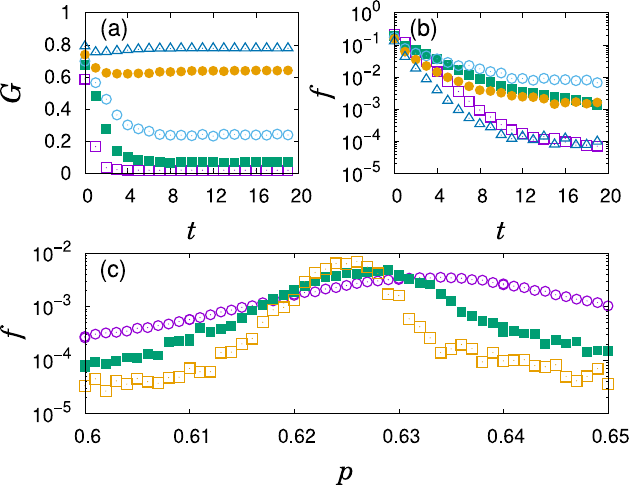}
\caption{(a) $G(t)$ for $N/10^3=16$ at $p=0.61$ $(\square)$, $0.62$ $(\blacksquare)$, $p_c$ $(\circ)$, $0.63$ $(\bullet)$, and $0.64$ $(\triangle)$, shown from bottom to top.
(b) $f(t)$ for $N/10^3=16$ at $p=0.61$ $(\square)$, $0.62$ $(\blacksquare)$, $p_c$ $(\circ)$, $0.63$ $(\bullet)$, and $0.64$ $(\triangle)$. (c) $f(p)$ for $N/10^3=1$ $(\circ)$,
$4$ $(\blacksquare)$, and $16$ $(\square)$.}
\label{Fig:Bipartite_GnNotordered}
\end{figure}

The order parameter is given by $G(p)$, which is the fraction of nodes belonging to the largest cluster at a given $p$.
$G(p)$ evolves as the number of rewirings per node, denoted by $t$, increases.
As shown in Fig.~\ref{Fig:Bipartite_GnNotordered}(a), $t$ increases until a steady state is reached, in which $G(p)$ fluctuates around a constant value. At the steady state, $G(p)$ exhibits a DPT at $p=p_c$, such that $G(p)>0(=0)$ for $p \geq p_c$ $(<p_c)$. The analytical derivation of $G(p)$ is summarized in Appendix A.

We note that the bipartite network is not a tree, and a steady state in which every node connects to neighboring clusters in ascending order of their sizes cannot, in general, be reached due to the presence of long range loops. 
In Fig.~\ref{Fig:Bipartite_GnNotordered}(b), we measure $f$ as a function of $t$, where $f$ denotes the fraction of nodes in the first partition that fail to connect to neighboring clusters in ascending order of their sizes.
We find that $f$ appears to converge to a finite value as $t$ increases, which supports that such a steady state is not reached.
In Fig.~\ref{Fig:Bipartite_GnNotordered}(c), the peak values of $f$ increase with increasing $N$, which further supports this expectation.
A clarification of why a tree network guarantees the existence of such a steady state, whereas a non-tree network does not, is provided using our dynamical EP process introduced below.

We now revise the model into a dynamical EP process where nodes rewire their links to reach a steady state after each link addition. Specifically, the involved nodes rewire once such that the largest cluster size reaches a steady state after a link is added. In this manner, the number of rewiring nodes remains finite below the threshold but diverges as the threshold is approached from below.
We note that the total number of rewiring nodes for each value of $p$ is extensive to $N$ 
as from the previous study~\cite{yschodiscPTtree2025}, as the rewiring
process begins from a randomly occupied network
that does not correspond to a steady state.

\begin{figure}[t!]
\includegraphics[width=0.8\linewidth]{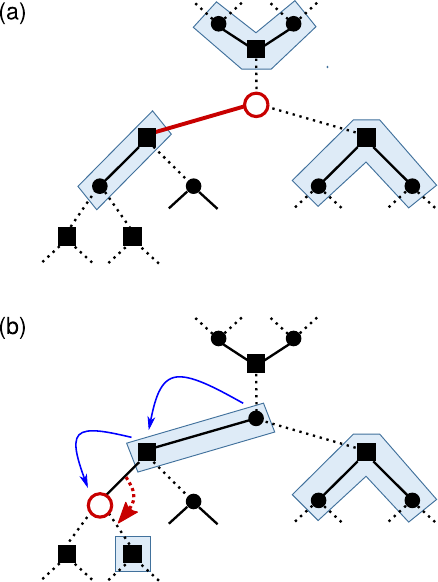}
\caption{Schematic diagram of link addition in the EP model on a bipartite network, where the circles (squares) represent nodes in the first (second) partition, solid lines denote occupied links, and dotted lines denote unoccupied links.
(a) One link (red line) is added to the initial rewiring node (empty circle) following steps (d1)--(d3).
(b) Following the solid arrows according to step (d4), the empty circle becomes the current rewiring node set. Then, it rewires according to steps (d2)--(d4). Since the next rewiring node set becomes empty in step (d5), the process terminates.
(a, b) We note that the two partitions alternate in the outward direction along the tree, since the bipartite network is locally tree-like, meaning that it appears as a tree within a finite distance from a given node.}
\label{Fig:SchematicBipartite}
\end{figure}

We introduce the dynamical rule of the EP model on the bipartite network as follows. At each time step $p \rightarrow p + 1/(N-1)$, one link is added according to the following steps:
\begin{itemize}
\item[(d1)] One unoccupied link is randomly selected and occupied. The node in the first partition at one end of this link becomes the initial rewiring node.
\item[(d2)] Each node in the current rewiring node set sequentially adds nodes to the next rewiring node set, following steps (d3) and (d4).
\item[(d3)] Each node rewires its links in ascending order of cluster sizes according to steps (c1)--(c3).
\item[(d4)] Every next neighbor (two steps distant) that is connected via occupied links from the rewiring node in (d3) and has never rewired before is added to the next rewiring node set.
\item[(d5)] If the next rewiring node set is empty, the process terminates. Otherwise, step (d2) is repeated, with the next rewiring node set becoming the current rewiring node set.
\end{itemize}
In short, rewiring nodes propagate outward from the initial rewiring node where the new link is added, and the propagation terminates once no further nodes are allowed to rewire.
A schematic diagram of this process is shown in Fig.~\ref{Fig:SchematicBipartite}.

\begin{figure}[t!]
\includegraphics[width=0.8\linewidth]{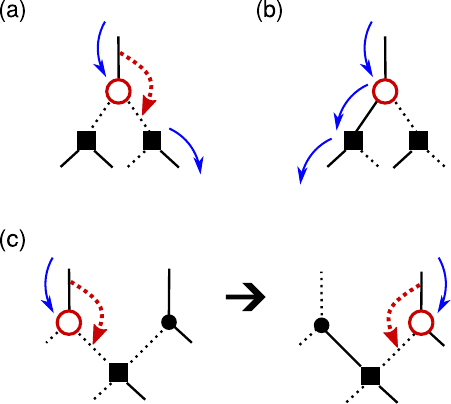}
\caption{Schematic diagram of the transient process following steps (d2)--(d4). (a) The circular node, approached via the upper arrow, rewires its link along the dotted arrow. The node reached via the bottom arrow becomes the next rewiring node. (b) The circular node reached via the upper arrow maintains its occupied links, and the node reached via the bottom arrow becomes the next rewiring node. (c) In the left panel, the circular node on the left rewires its link along the dotted arrow. In the right panel, the circular node on the right rewires its link to the common square node along the dotted arrow. }
\label{Fig:SchematicSteadyTree}
\end{figure}

We demonstrate that, in a tree network, all nodes remain ordered $(f=0)$ when a single link is added and the rewiring process progresses according to steps (d1)--(d5). For this purpose, the initial rewiring node in step (d1) is called the seed.
On the tree network, which includes both occupied and unoccupied links, the inner (outer) neighbor of a node is defined as a neighbor close to (far from) the seed.
Accordingly, the inner (outer) neighboring cluster of a node is the cluster that includes its inner (outer) neighbor, after the node disconnects all of its occupied links.

In step (d1), a single link is added to the seed in the tree network where all nodes are ordered.
We consider a node in the current rewiring node set in steps (d2)--(d4).

Before the node rewires, it should be connected to its inner neighbor, since it was reached from a node in the previous rewiring node set through that neighbor.
Here, the size of the inner neighboring cluster may have increased due to the addition of the link in step (d1).
If the increased size exceeds that of an outer neighboring cluster and the node is not connected to it, then the node rewires its inner link to the outer neighboring cluster, as shown in Fig.~\ref{Fig:SchematicSteadyTree}(a). Otherwise, it maintains its inner link, as shown in Fig.~\ref{Fig:SchematicSteadyTree}(b).
We note that the node remains ordered after rewiring. Moreover, it cannot rewire an outer link to the inner neighbor, as the inner neighbor is already connected prior to the rewiring.

After the node rewires, the sizes of its neighboring clusters may change as the rewiring process continues through repeated applications of steps (d2)--(d4).
It is apparent that the size of a disconnected outer neighboring cluster does not change, as no further process occurs outward through that neighbor. In contrast, the size of a connected neighboring cluster decreases monotonically as the rewiring process continues, since each rewiring node either maintains its inner link or rewires it in the outward direction, potentially disconnecting its branch from the cluster.
Therefore, the node remains ordered throughout the entire rewiring process until it terminates at step (d5).

In conclusion, if the network is a tree, all nodes remain ordered as the rewiring process progresses and eventually terminates after a single link is added in the EP model.
If the network is not a tree, then after a node rewires its inner link to an outer neighbor, another node may also connect to the same neighbor, as shown in Fig.~\ref{Fig:SchematicSteadyTree}(c). In this case, the size of the connected neighboring cluster increases, which may break the ordered state of the node after it has rewired. Under the dynamical rules (d1)--(d5), the node does not have an opportunity to rewire again, and thus a fraction of nodes may remain unordered when the process terminates.

\begin{figure}[t!]
\includegraphics[width=0.8\linewidth]{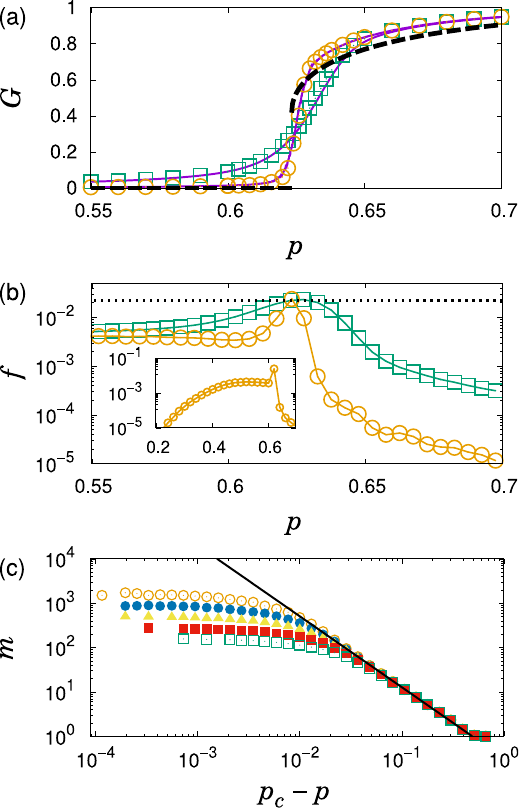}
\caption{
(a) $G(p)$ of the EP model (solid line) and the steady state (symbols) on the bipartite network with $N/10^3=1$ ($\square$ with line) and $16$ ($\bigcirc$ with line).     
The dashed line is the theoretical curve of $G(p)$
in the thermodynamic limit $N \rightarrow \infty$, described in Appendix A.
(b) $f(p)$ of the EP model near $p_c$ for $N/10^3=1$ and $16$ from bottom to top. The horizontal dotted line at $0.022$ serves as a guide to the maximum values of $f(p)$. Inset: $f(p)$ for $N/10^3=16$ over a wide range of $p$. (c) $m$ vs. $p_c-p$ on the bipartite network
for $N/10^3=1,2,4,8$, and $16$ from bottom to top. The slope of the solid line is $-8/5$. }
\label{Fig:BipartitePinf_nRewire}
\end{figure}

As shown in Fig.~\ref{Fig:BipartitePinf_nRewire}(a), $G(p)$ of the EP model coincides with that of the steady state from Appendix A, confirming that the EP model indeed exhibits a DPT.
This indicates that the EP model effectively suppresses the growth of the largest cluster, in a manner similar to the consistent rewiring of each node leading up to the steady state.
In Fig. \ref{Fig:BipartitePinf_nRewire}(b), we observe $f>0$ over a wide range of $p$ including $p_c$, as expected due to the fact that the bipartite network is not a tree. However, $f$ is small throughout this range, indicating that most rewiring nodes are ordered.
In Fig.~\ref{Fig:BipartitePinf_nRewire}(c), we measure $m$ for $p<p_c$ and find that $m \propto (p_c-p)^{-8/5}$, which supports that $m$ is finite for $p<p_c$ but diverges as $p_c$ is approached from below. As $m$ is finite for $p<p_c$, we assume that the rewiring nodes following the addition of a link form a tree structure, such that no two of them share a common outer neighbor during the propagation.
This assumption may explain why $f$ remains small and, accordingly, why $G(p)$ of the EP model coincides with that of the steady state.

One might ask whether such a small $f$ would make a difference between the $G$ of EP and steady state. To address this, we measure the fraction of nodes that do not satisfy a weak condition for the ordered state, where the weak condition requires that a node preferentially connect to neighbors not belonging to the largest cluster.
It is sufficient for each node to satisfy this condition in order to reveal a DPT by collaboratively suppressing the emergence of a macroscopic cluster. 
In Appendix B, we show that this fraction decreases with increasing system size in both the EP model and the steady state, suggesting that it vanishes in the thermodynamic limit $N \rightarrow \infty$.
Therefore, we conclude that the EP model exhibits a DPT by effectively inducing the steady state for each value of $p$.

In this result using the bipartite network, $m$ is finite for $p<p_c$ but diverges as $p_c$ is approached from below, in contrast to the branch (Sec.~\ref{sec:branch}) where $m$ remains finite up to $p_c$.
This is because the fraction of nodes in the macroscopic cluster at $p_c$ is finite in the bipartite network, whereas it vanishes to zero in the branch in the thermodynamic limit $N \rightarrow \infty$.
Thus, to effectively suppress the formation of a macroscopic cluster in the bipartite network, a divergent number of nodes must participate via local link rewiring as $p_c$ is approached from below.  
This may be related to the divergence of the average cluster size as $p_c$ is approached from below in typical percolation models~\cite{stauffer, kim_percolation}. 
A more detailed understanding of the behavior of $m$ is left for future work.

\section{Discussion: Distinction from Previous Explosive Percolation Models}
\label{sec:discussion}

In this section, we discuss in detail how this result differs from the previous EP models~\cite{explosive_phenomena}.
In previous EP models, at each time step an unoccupied link is occupied and remains occupied thereafter. In contrast, in our EP model, an occupied link may become unoccupied through link rewiring. Nevertheless, our EP model also satisfies the condition that the number of occupied links increases by one at each time step, without any cascade of link removals~\cite{InterdependentDorogovtsev, kcore3}. For this reason, our EP model can be regarded as a variation of the previous EP models~\cite{Achlioptas:2009}.

Moreover, in previous EP models exhibiting a DPT, information such as the sizes of attached clusters~\cite{Half_restricted} or the degrees of attached nodes~\cite{local2018} is used over an infinite number of link candidates at each time step. Conversely, our EP model on the branch requires information from only a finite number of links, yet it still exhibits a DPT. Our study therefore provides the first observation of a DPT in an EP model that relies on information from only a finite number of link candidates.

In a previous study~\cite{ziff_ncomm}, it was reported that random occupation on a one-dimensional lattice with long-range bonds also yields a DPT. We note, though, that random occupation is not an EP model, and the observed DPT in that case is driven by the underlying network structure rather than by suppression based on link candidates.

Our EP model on the bipartite network also requires information from a finite number of links for $p<p_c$. This differs from previous EP models exhibiting a DPT that require information from an infinite number of links for $p<p_c$~\cite{riordan}. In conclusion, our study demonstrates that the introduction of link rewiring enables an EP model to exhibit a DPT while relying solely on local information.

\section*{Acknowledgments}
This research was supported by the “Research Base Construction Fund Support Program” funded by Jeonbuk National University in 2024,
and also by the National Research Foundation (NRF) of Korea, grant no. RS-2025-16067164.

\section*{Data availability}
Codes and data are available at https://github.com/youngsulcho/DPT\_in\_LocalEP.git.
\section*{Appendix A: Analytical solution of the order parameter}

We provide the analytical solution of $P_{\infty}(p)$ in the thermodynamic limit $N \rightarrow \infty$ on the branch with $z-1=3$ (dashed line in Fig.~\ref{Fig:BranchPinf_FracRewire}(a)); the detailed derivation is presented in~\cite{yschosciadv2025}. For a given $p$, each node occupies $n$ $(0 \leq n \leq 3)$ links with probability $Q(n,p)=\binom{3}{n}p^n(1-p)^{3-n}$. In the steady state, each node $i$ connects its $n$ links to neighbors in ascending order of their cluster sizes. For node $i$ to be connected to a macroscopic cluster, at least $4-n$ of its neighbors must belong to a macroscopic cluster, because node $i$ preferentially connects its links to finite neighboring clusters. Each neighbor independently belongs to a macroscopic cluster with probability $P_{\infty}(p)$, and the probability that node $i$ belongs to a macroscopic cluster should be $P_{\infty}(p)$ self-consistently. As a result, $P_{\infty}(p)$ satisfies the equation
\begin{align}
P_{\infty}(p)=&Q(1,p)P_{\infty}^3(p)+Q(2,p)(3P_{\infty}^2(p)-2P_{\infty}^3(p)) \notag \\
           &+Q(3,p)[1-(1-P_{\infty}(p))^3].
\tag{A1}
\label{eq:branchPinf}
\end{align}
The solution of Eq.~(\ref{eq:branchPinf}) yields a discontinuity of $P_{\infty}(p_c) \approx 0.665286$ at $p_c \approx 0.652834$, as described by the dashed line in Fig.~\ref{Fig:BranchPinf_FracRewire}(a).

Next, we provide the analytical solution of $G(p)$ in the thermodynamic limit $N \rightarrow \infty$ for the bipartite network (dashed line in Fig.~\ref{Fig:BipartitePinf_nRewire}(a)); the detailed derivation is presented in~\cite{yschodiscPTtree2025}. For a given $p$, we introduce two probabilities, $G_1(p)$ and $G_2(p)$, defined at the steady state. Here, $G_1(p)$ represents the probability that a node in the first partition belongs to a macroscopic cluster, while $G_2(p)$ represents the probability that its neighbor is connected to a macroscopic cluster via outgoing links. Then, the self-consistency equation for $G_1(p)$ is obtained by applying the suppression rule and replacing $P_{\infty}(p)$ with $G_2(p)$ on the right-hand side of Eq.~(\ref{eq:branchPinf}), yielding
\begin{align}
G_1(p)=&Q(1,p)G_2^3(p)+Q(2,p)[3G_2^2(p)-2G_2^3(p)] \notag \\
           &+Q(3,p)[1-(1-G_2(p))^3]
\tag{A2}
\label{eq:bipartitePinf}
\end{align}
with the relation $G_2(p)=1-(1-pG_1(p))^2$.
By numerically solving Eq.~(\ref{eq:bipartitePinf}) to obtain $G_1(p)$ for each $p$, we obtain the order parameter $G(p)=[G_1(p)+1-(1-pG_1(p))^3]/2$, which is shown as the dashed line in Fig.~\ref{Fig:BipartitePinf_nRewire}(a) along with the discontinuity of $G(p_c) \approx 0.417456$ at $p_c \approx 0.623339$.

\section*{Appendix B: Suppression of a Macroscopic Cluster by all Nodes in the Bipartite Network}

In each network configuration, the size of the largest clusters is uniquely determined, and one of the clusters of that size is designated as the largest cluster. A sufficient condition for a DPT is that every node avoids connecting to its neighbors belonging to the largest cluster. In other words, each node first connects its links to neighbors outside the largest cluster, and only then connects any remaining links to neighbors within the largest cluster, if such neighbors exist. In this manner, the emergence of a macroscopic cluster is suppressed by a finite fraction of nodes that would otherwise be connected to neighbors in the macroscopic cluster. If the sufficient condition is satisfied, a DPT emerges in accordance with the analytic result of $G(p)$ (dashed line in Fig.~\ref{Fig:BipartitePinf_nRewire}(a)).

\begin{figure}[t!]
\includegraphics[width=1.0\linewidth]{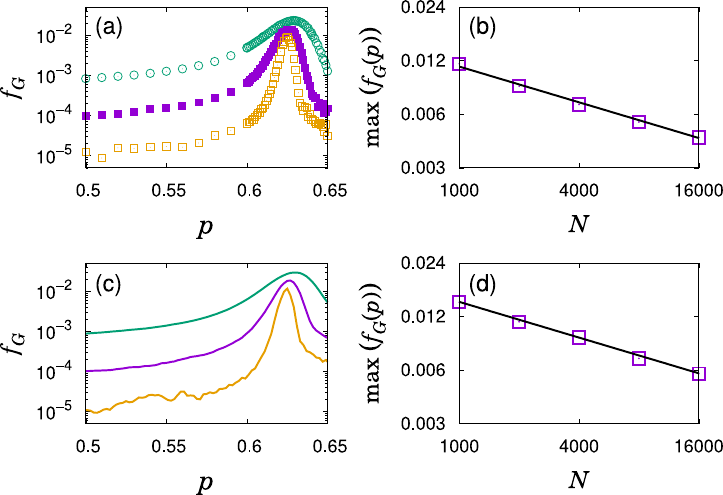}
\caption{
(a) $f_G(p)$ in the steady state of the bipartite network for $N/10^3 = 1$ $(\circ)$, $4$ $(\blacksquare)$, and $16$ $(\square)$, shown from top to bottom. (b) Maximum of $f_G(p)$ from (a) as a function of $N$. The slope of the solid line is $-1/3$. (c) $f_G(p)$ of the EP model in the bipartite network for $N/10^3 = 1$, $4$, and $16$ from top to bottom. (d) Maximum of $f_G(p)$ from (c) as a function of $N$. The slope of the solid line is $-1/3$.}
\label{Fig:Bipartite_nNotorderedInf}
\end{figure}

We examine whether the sufficient condition holds in both the steady state and the EP model. To this end, we measure $f_G(p)$, defined as the fraction of nodes that belong to the largest cluster and have at least one disconnected neighbor outside the largest cluster at a given $p$. In Fig.~\ref{Fig:Bipartite_nNotorderedInf}(a) and (b), we find that $f_G(p)$ rapidly decreases with increasing $N$ over the entire range of $p$, and that the maximum value of $f_G(p)$ decreases as $\text{max} (f_G(p)) \propto N^{-1/3}$ in the steady state. 
In Fig.~\ref{Fig:Bipartite_nNotorderedInf}(c) and (d), we observe similar results for the EP model, supporting that the sufficient condition holds in both the steady state and the EP model.

We briefly explain why $f_G(p)=0$ in the thermodynamic limit, in contrast to $f(p)>0$. In the right panel of Fig.~\ref{Fig:SchematicSteadyTree}(c), the circle node on the right would otherwise be connected to the largest cluster via the upper link, but it becomes disconnected due to link rewiring. As a result, although the circle node on the left may become unordered leading to an increase in $f$, it does not connect to the largest cluster through this process, and thus $f_G$ does not increase. This difference between $f$ and $f_G$ explains why $f_G(p)=0$ in the thermodynamic limit even in non-tree networks.

\bibliography{references.bib}

\end{document}